\documentclass[useAMS,usenatbib]{mn2e}
\usepackage{graphicx}
\usepackage{amsmath,amssymb}

\newcommand{\BF}{\begin{figure}\begin{center}}
\newcommand{\EF}{\end{center}\end{figure}}
\newcommand{\BE}{\begin{equation}}
\newcommand{\EE}{\end{equation}}
\newcommand{\BEA}{\begin{eqnarray}}
\newcommand{\EEA}{\end{eqnarray}}

\newcommand{\tr}{\textrm}

\newcommand{\bvec}[1]{\mbox{\boldmath $#1$}}

\newcommand{\ms}{M_{\odot}}
\begin{document}
\title[Weak Lensing by Minifilament or Minivoid as the Origin of Flux-ratio
Anomalies]{Weak Lensing by Minifilament or Minivoid as the Origin of Flux-ratio
Anomalies in Lensed Quasar MG0414+0534}
\author[Kaiki Taro Inoue]
{Kaiki Taro Inoue$^1$\thanks{E-mail:kinoue@phys.kindai.ac.jp}
\\
$^{1}$Department of Science and Engineering, 
Kinki University, Higashi-Osaka, 577-8502, Japan  }

\date{\today}

\pagerange{\pageref{firstpage}--\pageref{lastpage}} \pubyear{0000}

\maketitle

\label{firstpage}
\begin{abstract}
We explore the weak lensing effects by ministructures
in the line-of-sight in a quadruply lensed quasar MG0414+0534 
that shows an anomaly in the flux ratios. 
We find that the observed flux-ratio anomaly can be explained
by a presence of either a minifilament or a minivoid in the
 line-of-sight with a 
surface mass density of the order of $10^{8-9}\,h^{-1}\ms/\tr{arcsec}^2$ without taking into
account any subhalos in the lensing galaxy. The astrometric perturbation  
by a possible minifilament/minivoid at the primary lens plane is $\lesssim
 0.001\,\tr{arcsec}$ and the amplitudes of convergence perturbations 
due to these perturbers are $\delta \kappa \sim 0.004-0.008$ at the place of an 
image that shows anomaly. In order to discriminate
models with the line-of-sight ministructures 
from those with a subhalo(s) in the lensing galaxy, 
we need to precisely measure the projected convergence and shear perturbations
around the lensing galaxy. The differential magnification effect could break the
model degeneracy if the source size is $ \gtrsim 100\,\tr{pc}$. 
Observation at the submillimeter band using 
interferometers will enable us to
determine the origin of anomalies in the flux ratios. 
\end{abstract}

\section{Introduction}
The physics of non-linear clustering on scales of $\lesssim 1\,$Mpc
has not been well understood.
First, the observed 
density profiles of dwarf galaxies are shallower than
the values predicted from $N$-body simulations, which is called the
'cusp-core problem' \citep{navarro1996, moore1999, swaters2003,
simon2005}. Second, the number of observed satellite galaxies in our Galaxy is 
significantly smaller than the predicted value, which is known as
the 'missing  satellites problem' \citep{klypin1999,moore1999}. Third, the observed circular
velocities of most massive subhalos in our Galaxy are also significantly
smaller than the predicted values, which is called the 'too big to fail 
problem' \citep{boylan-kolchin2011, wang2012}.   

In order to address these issues, it is important 
to constrain the number density of
dark satellites in extragalactic halos. It has been known that 
some quadruply lensed quasars show anomalies in the observed 
flux ratios of lensed images provided that 
the gravitational potential of the lens is
sufficiently smooth. Such a discrepancy is 
called the ``anomalous flux ratio'' and 
has been considered as an imprint of cold dark matter (CDM) subhalos
with a mass of $\sim 10^{8-9} \ms$ in lensing
galaxies \citep{mao1998,metcalf2001,chiba2002,dalal-kochanek2002,
keeton2003, metcalf2004,chiba2005,sugai2007,mckean2007,
more2009,minezaki2009, xu2009,xu2010}. 

However, intergalactic halos in the line-of-sight 
may act as perturbers as well\citep{chen2003,metcalf2005a,xu2012}.
Indeed, taking into account the astrometric shifts, recent studies 
have found that the observed anomalous flux ratios can be 
explained solely by these line-of-sight structures with a surface density $\sim 
10^{7-8}\, h^{-1}\ms/\textrm{arcsec}^2$ \citep{inoue-takahashi2012,
takahashi-inoue2014,inoue-etal2014} without taking into account 
subhalos in the lensing galaxies. The observed increase in the amplitude of  
magnification perturbations as a function of the source redshift 
strongly implies that the origin is associated with sources rather than lenses.

In order to determine the origin of flux-ratio anomalies, we need to 
precisely measure the perturbation of 
gravitational potential projected on a plane 
perpendicular to the line-of-sight. If it is caused by a subhalo in the lensing galaxy halo, 
then the perturbation effect is limited to the region around the
perturber. Even if the perturbation 
is caused by a number of subhalos in the lensing galaxy, 
we may be able to neglect 
the spatial correlation of between perturbers, as 
structures such as filaments and walls can be easily 
destroyed due to tidal force in the potential of the parent 
halo\footnote{If tidal streams in galaxy halos persist for long time, they
might mimic the lensing effects by filaments in the intergalactic space. }.

On the other hand, if the perturbation is caused by objects in the
line-of-sight, the spatial correlation between perturbers 
may be important than previously thought. 
In the CDM scenarios, the cosmic web structures appear on all the scales
that exceed the free-streaming scale. Therefore, any clumps formed on walls
and filaments should have spatial correlations between them. Moreover, 
inter-galactic medium may reside along these structures and
enhance the lensing effect by these clumps due to radiative cooling. In fact,
recent studies on Lyman-$\alpha$ emission and high-velocity clouds around the Milky
Way suggest an existence of substantial cold gas 
accreted on filaments (on scales $<$1Mpc) of the ancient cosmic web
\citep{dekel2009, cantalupo2014, martin2014}. These spatially extended
objects (filaments or walls) can perturb the fluxes of lensed images
if they are common in the universe. 

In the following, we call them ``ministructures'' if the typical
length-scale of these structures (dark matter + baryon) is just $\sim 
10-100 \,$kpc \citep{inoue-takahashi2012}, which is 
significantly smaller than those (on scales of $10-100\,$Mpc) 
usually discussed in literature \citep{colberg2005, mead2010, 
higuchi2014}. 

In this paper, we investigate whether the flux-ratio anomaly
in an anomalous quadruple lens MG0414+0534 
can be explained by a minivoid or a minifilament in the line-of-sight
and study whether we can break the model degeneracy. 
For simplicity, we put a perturber at the primary 
lens plane where the deflection angle is relatively large
\citep{inoue-takahashi2012}. Although, the objects in the line-of-sight may have
complex structures, we use simple toy models to arrive at some quick and
dirty results. Since the gravitational effects of 
non-linear ministructures are different from subhalos, we may be 
able to distinguish between them by measuring the differential
magnification effect provided that the source size is sufficiently large.
In section 2, we present the observational data of 
MG0414+0534. In section 3, we describe a simple lens model of a
tidally truncated singular isothermal sphere (SIS), 
a compensated homogeneous filament and void.
In section 4, we show our results of 
$\chi^2$ fitting and the gravitational effect 
in models with a minifilament/minivoid.
In section 5, the differential magnification effect 
in minifilament/minivoid models is discussed.
In section 6, we summarize our result 
and discuss some relevant issues and future prospects.

In what follows, we assume a cosmology 
with a matter density $\Omega_{m,0}=0.3134$, a baryon density 
$\Omega_{b,0}=0.0487$, a cosmological constant $\Omega_{\Lambda,0}=0.6866$,
a Hubble constant $H_0=67.3\, \textrm{km}/\textrm{s}/\textrm{Mpc}$,
a spectral index $n_s=0.9603$, and the root-mean-square (rms) 
amplitude of matter fluctuations at $8 h^{-1}\, \textrm{Mpc}$, 
$\sigma_8=0.8421$, which are obtained from the observed 
CMB (Planck+WMAP polarization; \citet{ade2013}). $G$ and $c$ denote
the gravitational constant and speed of light, respectively.
\section{MG0414+0534}
\begin{figure}
\begin{minipage}{0.49\textwidth}
\hspace{0.2cm}
\includegraphics[width=82mm]{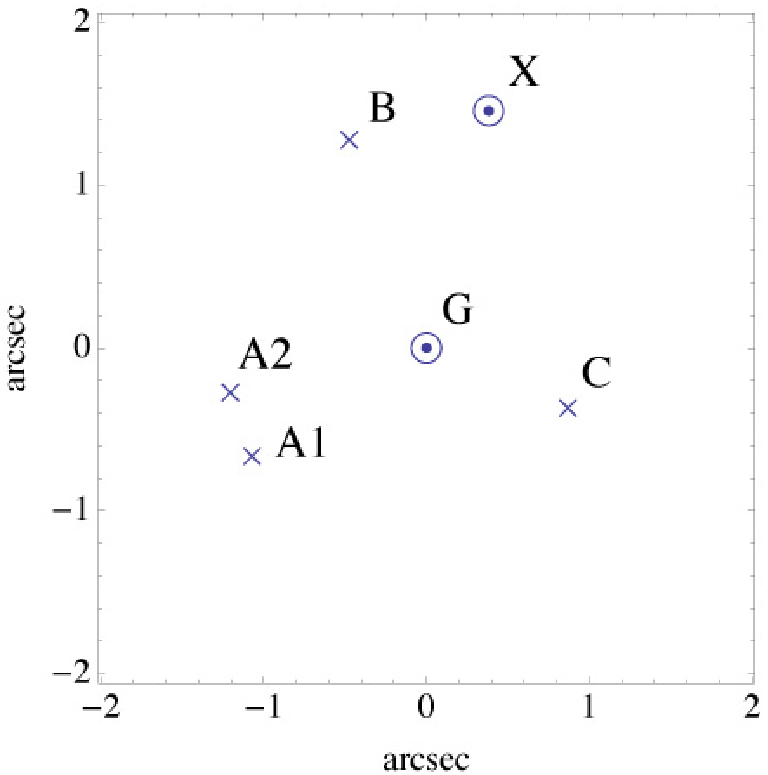}
\caption{The gravitational lens system MG0414+0534 (plots of data in
 table 1).~~~~~~~~~~}
\label{mg0414}
\end{minipage}
\vspace*{0.6cm}
\begin{minipage}{0.5\textwidth}
\makeatletter
\def\@captype{table}
\makeatother
\caption{The observed HST positions of lensed images (A1,A2,B,C) 
and the centroids of the primary lens G and object X and the 
MIR flux ratios in MG0414+0534.~~}
\setlength{\tabcolsep}{4pt}
\vspace{-0.1cm}
\begin{tabular}{lcc}
\hline
\hline
 Image & Position(arcsec) & MIR flux ratio(obs.)  \\ 
\hline
  A1 & $(-1.07\pm0.003,-0.667\pm0.003)$ & \\ 
   A2 & $(-1.203\pm0.003,-0.274\pm0.003)$ & A2/A1=$0.919 \pm 0.021$ \\
  B & $(-0.474\pm 0.000,1.276\pm 0.000)$ & B/A1=$0.347 \pm 0.013$ \\
   C & $(0.871\pm0.003,-0.372\pm0.003)$ & \\
   G & $(0.000 \pm0.003,0.000\pm0.003)$ & \\
   X & $(0.383\pm0.011,1.457\pm0.009)$ & \\
\hline
\label{table1}
\end{tabular}
\\ Note: The positions are taken from CASTLES data base
(\citet{falco1997}) and are relative to image G. 
The MIR flux ratios are taken from the combined results of 
\citet{minezaki2009} and \citet{macleod2013}. 
\end{minipage}
\end{figure}

The fold-caustic lens MG0414+0534 consists of two bright images A1, A2 and 
two faint images B, C. The images A1 and B are minima, and A2 and C are saddles.
A source quasar at $z_S=2.64$ is lensed by an elliptical galaxy
(hereinafter referred to as G) at
 $z_L=0.96$ \citep{hewitt1992, lawrence1995, tonry1999}.
A simple lens model, a singular isothermal ellipsoid (SIE)
\citep{kormann1994} with
an external shear (ES) cannot fit the image positions as well as 
the flux ratios. \citet{schechter1993} and \citet{ros2000} 
suggested that another galaxy is necessary for
fitting the relative image positions.
As shown in table \ref{table1}, we use the 
MIR flux ratios A2/A1 and B/A1 obtained from measurements 
by \citet{minezaki2009} and \citet{macleod2013}, since the 
radio fluxes might be hampered by Galactic refractive
scintillation\citep{koopmans2003}. For the astrometry, 
we use the data from CASTLES data base. However, in our analysis, 
we do not use the VLBI data \citep{ros2000,trotter2000} 
as there is an ambiguity in identifying 
components of jets and estimating the position errors 
due to large shear at the places of lensed images. 
 
Although, the observed image positions are well
fitted by an SIE and an external shear (SIE-ES) plus an SIS at the lens
redshift $z_l$, that accounts for object X, the 
flux ratios are not well fitted. A possible subhalo 
near A2 significantly improves the fit
\citep{minezaki2009,macleod2013}. The 
orientation of the external shear 
is in the same general direction as another 
object G6 found $4.^{''}4$ southwest of image C in the HST H-band image 
but it is not massive enough to fully account for the external shear
\citep{macleod2013}.

In this paper, we use a model that
consists of an SIE (for G) plus an
external shear (for environment), and an SIS (for X)
whose predicted positions of lensed images and centroids of galaxies 
are fitted to the observed data. We do not use 
the data of relative fluxes of lensed images for estimating the 
unperturbed model. The flux ratios are used for parameter
fitting when a perturber (subhalo/filament/void) is 
taken into account.
\section{Lens model}
For simplicity, as a perturber, we 
consider three types of models: a tidally truncated SIS halo, 
a compensated homogeneous filament and a compensated homogeneous void. 
Because of mass compensation, the gravitational potential of a perturber 
vanishes at the outer boundary in the latter two cases. Any locally 
uncompensated filaments (voids) can be modelled by adjusting the size of
the outer positive (negative) region.
\subsection{Subhalo}
As a simple model of subhalo in the primary lensing
galaxy, we consider a tidally truncated SIS (hereinafter referred to as
tSIS).  At a distance 
$R$ from the center of the primary lensing galactic
halo with a one-dimensional velocity dispersion $\sigma_0$, the tidal
radius $R_t$ (proper length) is approximately given by $R_t\approx R \sigma/\sigma_0$
where $\sigma$ is the one-dimensional velocity dispersion of a tSIS. 
In terms of the Einstein radius (mass scale) $b_h$ 
of a tSIS and that of an SIS for a parent
galaxy halo $b_0$, the 
tidal radius can be written as $R_t=R \sqrt{b_h/b_0}$. 

In what follows, we assume that the tSIS
resides at the lens plane for simplicity. In this case, $R$ coincides 
with the angular distance between the center of the primary lens and
that of a tSIS. In general, however, the tidal radius can be
larger than the value in which a tSIS resides at the lens plane for a given angular distance from
the center of the primary lens. The deflection angle $\hat{\bvec{\alpha}_h}$ is constant
inside but it decays as $\hat{\alpha_h} \propto 1/R $ outside the tidal radius.
\subsection{Compensated homogeneous filament}
We consider an infinitely long filament that 
has a positive constant density $\rho_+$ inside
a radius $R_+$ (region I) and a negative constant density $\rho_-<0$ at
$R_+<R<R_-$ (region II), where $R$ denotes the proper 
distance from the axis. At $R>R_-$ (region III), $\rho=0$ (Fig. 1). From
the compensating condition, we have 
\BE
\rho_-=-\frac{\rho_+R_+^2}{R_-^2-R_+^2}.
\EE
If we set the gravitational potential at region III equal to zero, i.e., 
$\psi_{III}(R)=0$, potentials at region I and II are
\BEA
\psi_I(R)&=&\pi G
\rho_+\biggl[(R^2-R_+^2)+\frac{2R_+^2}{R_-^2-R_+^2}\biggl(R_-^2\ln\frac{R_+}{R_-}
\nonumber
\\
&-&\frac{R_+^2-R_-^2}{2}
 \biggr)  \biggr],
\EEA
and
\BE
\psi_{II}(R)=\frac{2 \pi G \rho_+ R_+^2}{R_-^2-R_+^2}\biggl[R_-^2\ln\frac{R}{R_-}-\frac{R^2-R_-^2}{2}  \biggr].
\EE
Integrating the gravitational potential $\psi(\bvec{R}_\perp,Z)$ at
 $-\infty<Z<\infty$ where $\bvec{R}_\perp=(X,Y)$ are the orthogonal proper 
coordinates at the lens
plane and $z$ is the proper coordinate along the line-of-sight, and taking 
the gradient, we have the $X$-component of the 
deflection angle $\hat{\alpha}_f$ at $\bvec{R}_\perp=(X,Y)$,
\BE
\hat{\alpha}_f= b_f\times \left\{ 
\begin{array}{ll}
0, & \mbox{$R_- \le |\bvec{R}_\perp| $} \\
\dfrac{1}{R_-^2-R_+^2}
\biggl[ -X\sqrt{R_-^2-X^2}
\\ +R_-^2
\tan^{-1}\biggl(\dfrac{\sqrt{R_-^2-X^2}}{X}\biggr)

\biggr], & \mbox{$R_+< |\bvec{R}_\perp| < R_-$} \\

\dfrac{1}{R_-^2-R_+^2}
\biggl[ -X\sqrt{R_-^2-X^2}
\\
\\
+2X\sqrt{R_+^2-X^2}
\\
+R_-^2
\tan^{-1}\biggl(\dfrac{\sqrt{R_-^2-X^2}}{X}\biggr)
\\ -R_-^2
\tan^{-1}\biggl(\dfrac{\sqrt{R_+^2-X^2}}{X}\biggr)

\biggr], & \mbox{$0\le |\bvec{R}_\perp| < R_+$},
\end{array}
\right.
\EE
where $b_f$ describes the mass scale, which 
is given by
\BE
b_f \equiv \frac{8 \pi G R_+^2 \rho_+}{c^2}.
\EE 
In terms of linear density $\lambda$, $\theta_f$ can be also written as
\BE
b_f =\frac{8 G \lambda}{c^2}.
\EE
Because the filament is axi-symmetric, $Y$-component of the deflection
angle $\hat{\bvec{\alpha}}_f $ is zero. The deflection angle $\hat{\alpha}_f$
as a function of $X/R_+$ is shown in Fig. \ref{plots-alpha}.
\vspace*{1.0cm}
\begin{figure}
\includegraphics[width=41mm]{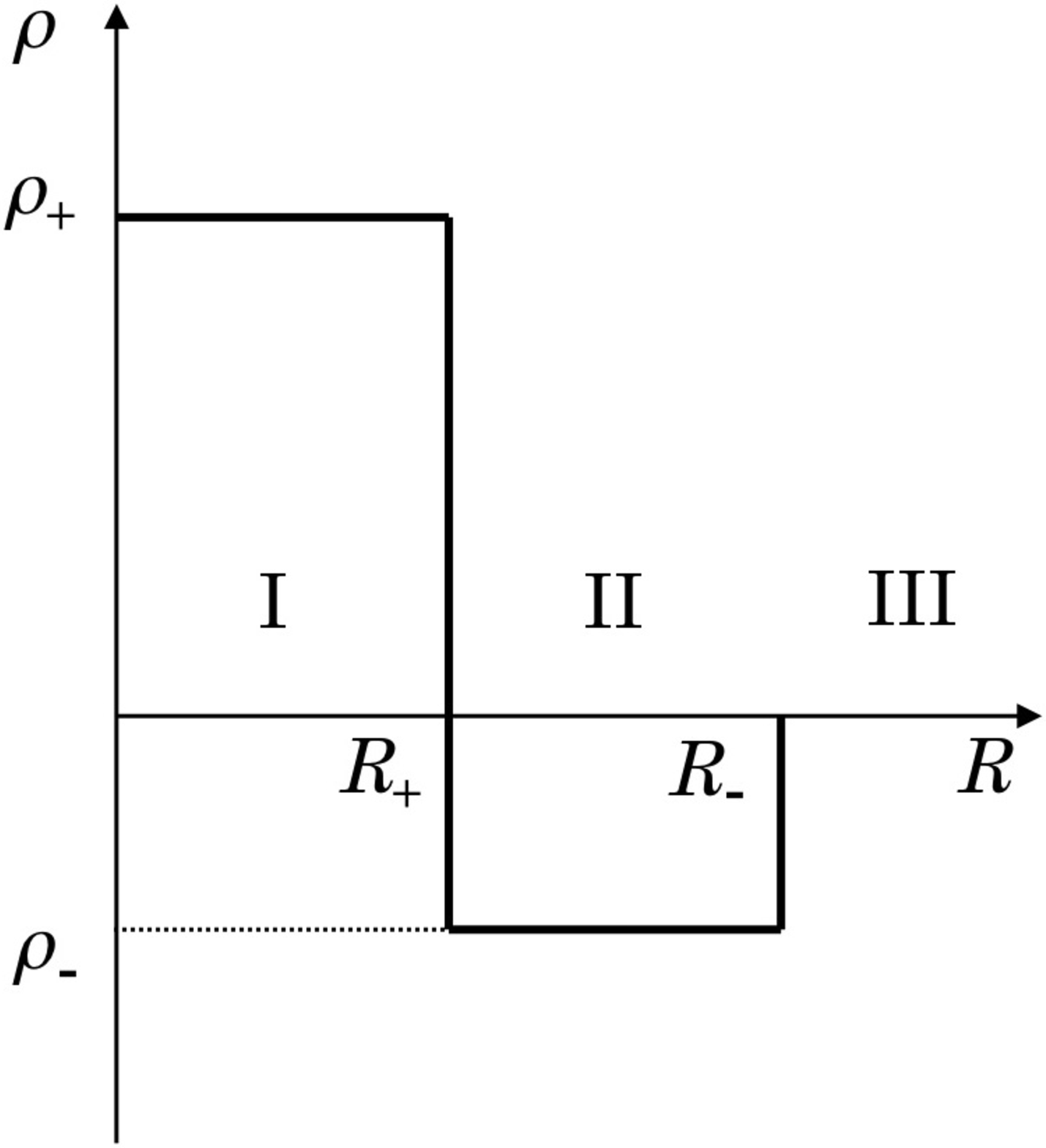}
\includegraphics[width=45mm]{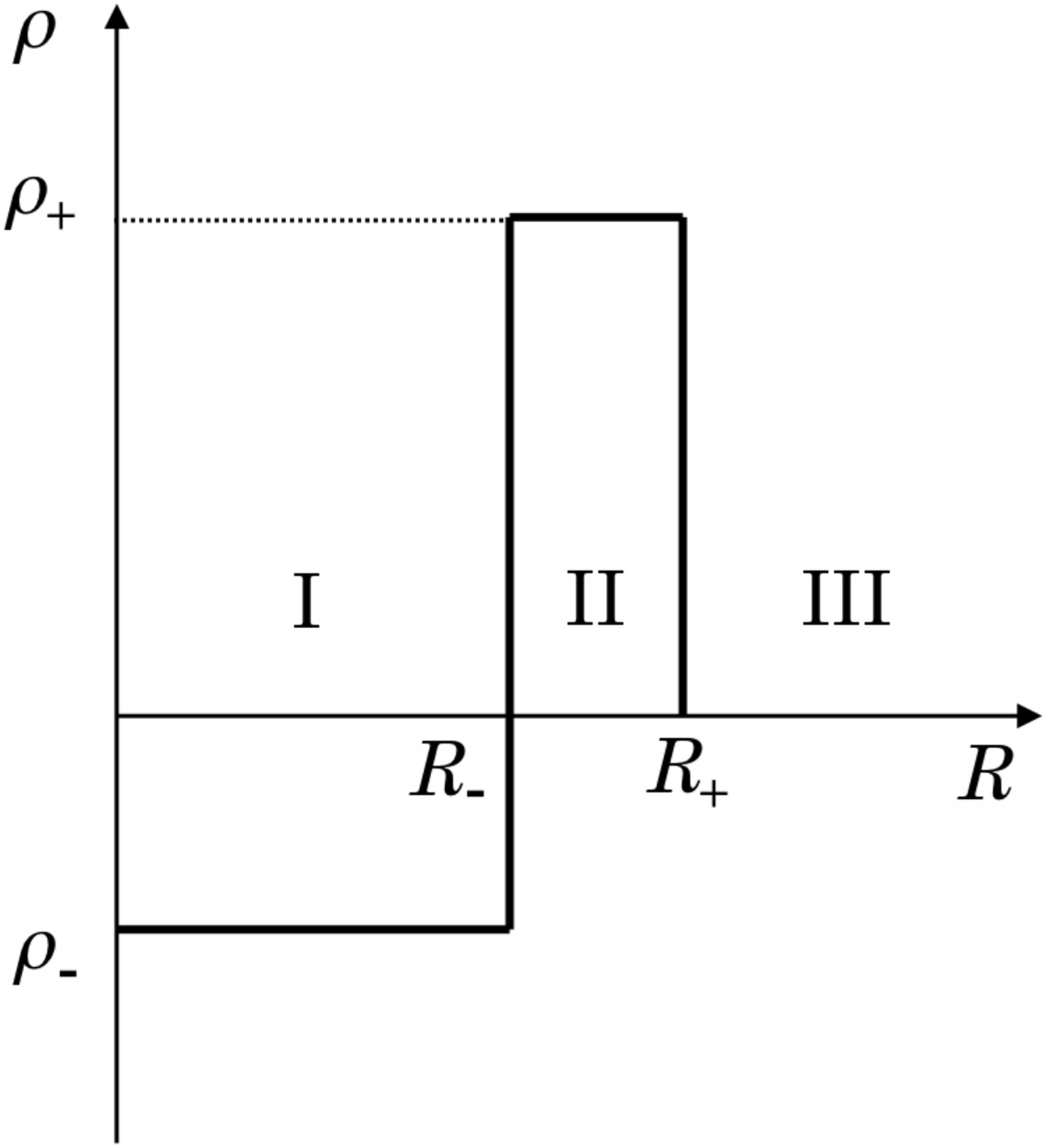}
\caption{Mass density of a compensated filament (left) and that of a compensated 
void (right). }
\label{density}
\vspace*{0.2cm}
\end{figure}
\vspace*{1.0cm}
\begin{figure}
\hspace{-0.3cm}
   \includegraphics[width=42mm]{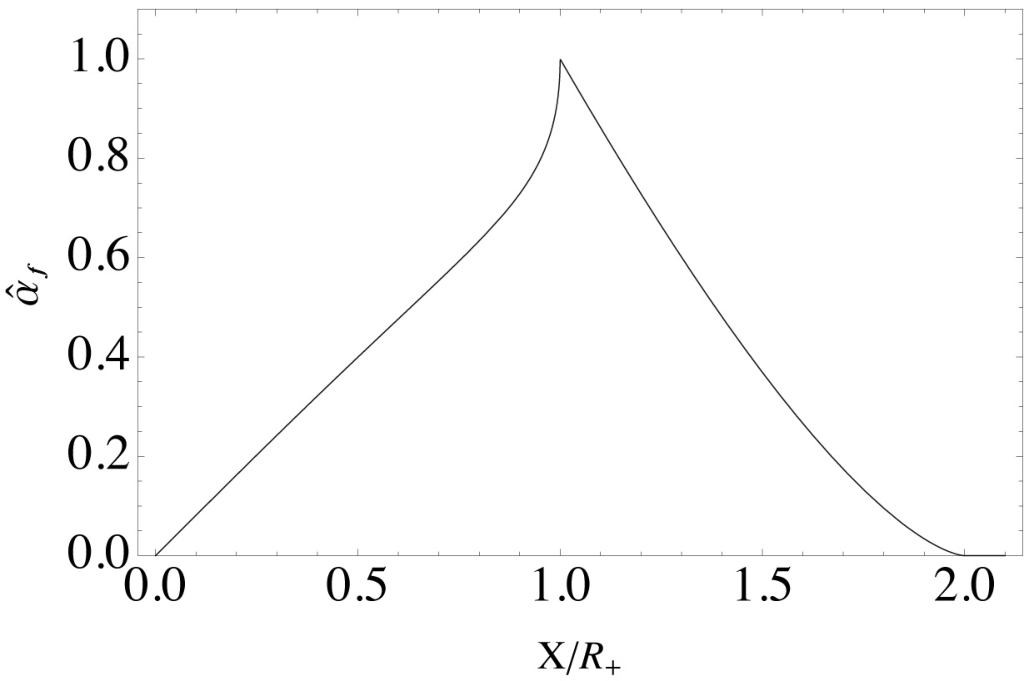}
\hspace{-0.1cm}
   \includegraphics[width=44mm]{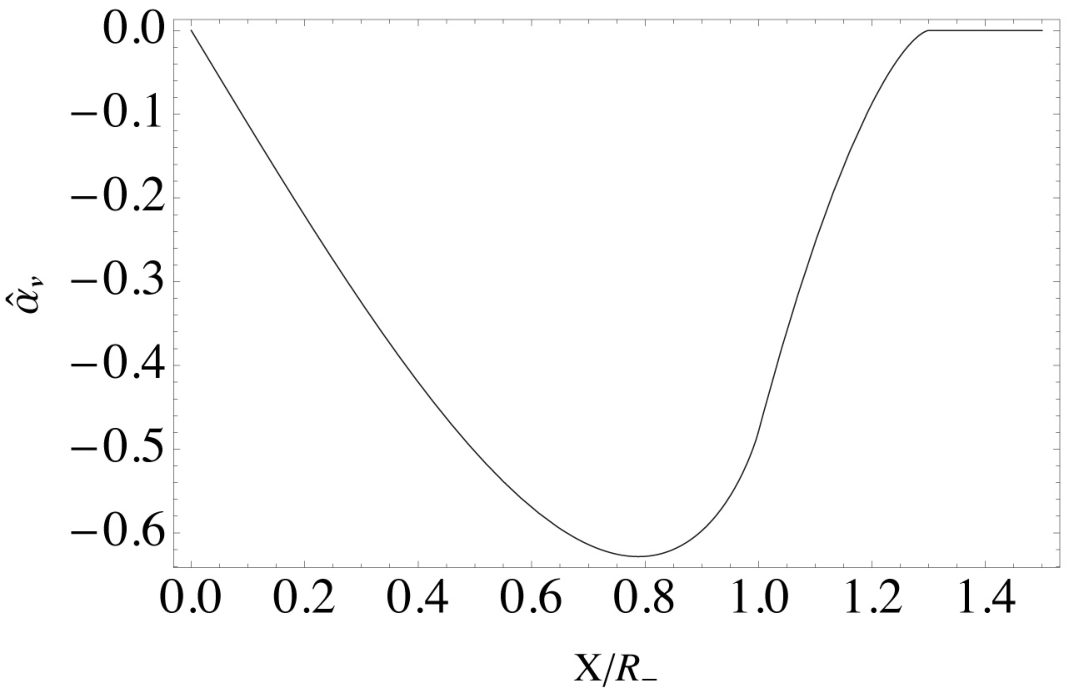}
  \caption{Deflection angle of a compensated filament (left) and that of a compensated 
void (right) }
  \label{plots-alpha}
\end{figure}
\subsection{Compensated homogeneous void}
For simplicity, we consider homogeneous spherical voids 
that are compensated in mass. We assume that 
they have a negative constant density $\rho_-$ inside
a radius $R_-$ (region I) and a positive constant density $\rho_+$ at
$R_+<R<R_-$ (region II), where $R$ denotes the proper 
distance from the center of a void. At $R>R_+$ (region III), the density
is vanishing, i.e., $\rho=0$ (Fig. 1). The mass deficient at  $R<R_-$ is
$M_v=4 \pi \rho_- R_-^3/3$. Then, the $X$-component of 
the deflection angle $\hat{\alpha}_v$ at 
$\bvec{R}_\perp=(X,Y=0)$ in the lens plane 
is \citep{amendola1999},
\BE
\hat{\alpha}_v= b_v\times \left\{ 
\begin{array}{ll}
0, & \mbox{$ R_+ \le |\bvec{R}_\perp|$} \\
\\
\tilde{X}^{-1}(\tilde{d}^3+3\tilde{d}^2+3\tilde{d})^{-1}
\\
\times \bigl[(1+\tilde{d})^2-\tilde{X}^2)\bigr]^{3/2}
, & \mbox{$R_-<|\bvec{R}_\perp|< R_+$} \\
\\
\tilde{X}^{-1}(\tilde{d}^3+3\tilde{d}^2+3\tilde{d})^{-1}
\\
\times 
\Bigl[ \bigl[(1+\tilde{d})^2-\tilde{X}^2)\bigr]^{3/2}
\\
-(1+\tilde{d})^3(1-\tilde{X}^2)^{3/2} \Bigr]
, & \mbox{$0\le |\bvec{R}_\perp| < R_-$},
\end{array}
\right.
\label{eq:void}
\EE
where $\tilde{X}\equiv X/R_-$, $\tilde{d}\equiv R_+/R_--1$, 
and $b_v$ describes the mass scale, which 
is given by
\BEA
~~~~~~~~~~~~~~~~~~~~~b_v &\equiv& \frac{4 G M_v }{c^2 R_-}
\nonumber
\\
&=& \frac{16 \pi G R_v^2 \rho_-}{3 c^2}.
\EEA
The $Y$-component of $\hat{\bvec{\alpha}}_v$ is zero for $Y=0$.
For $Y\ne 0$, a rotation of coordinates by an angle 
$\phi=\tan^{-1}(Y/X)$ gives the deflection angle 
$\hat{\bvec{\alpha}}_v=(\hat{\alpha}_v,0)$
as the void is spherically symmetric. 
The reduced deflection angle $\alpha_v$
as a function of $X/R_-$ is shown in Fig. \ref{plots-alpha}.

\section{Simulation}
For simplicity, we consider models
that consist of an SIE-ES (G) plus an SIS (X) perturbed by either a
tidally cut SIS, a compensated filament, or a compensated void. 
The parameters of the SIE-ES is the mass scale
$b_\tr{G}$\footnote{$b_{\tr{G}}$ corresponds to the 
Einstein radius $\theta_E$ of an SIS whose mass
inside $\theta_E$ is equal to the mass inside the critical curve of an
SIE \citep{kormann1994}.}, the position of 
the primary lens $(x_\tr{G}, y_\tr{G})$, 
the ellipticity $e(G)$, the external shear
$\gamma$, and their angles, $\phi_{e(G)}$, $\phi_\gamma$ measured in East of
North. The parameters of SIS 
is the Einstein radius (mass scale) 
$b_\tr{X}$ and the position $(x_\tr{X}, y_\tr{X})$. 
The parameters of perturbers are the positions $(x_h,y_h),(x_f,y_f),
(x_v,y_v)$ and the mass scales $b_h,b_f,b_v$ for the subhalo, filament
and void models, respectively. $(x_h,y_h)$ and $(x_v,y_v)$ correspond
to the centers of circular symmetry in a subhalo and a void, respectively. 
$(x_f,y_f)$ denotes the position on the axis of a filament that is closest   
to the center of the primary lens. The extent of filament and void
are specified by $R_+$ and $R_-$ (see section 3 for definition).
The position of a source is denoted as $(x_s,y_s)$.

As an unperturbed model, we adopt an SIE-ES plus an SIS 
and use only observed positions of lensed images and centroids of
lensing galaxies for fitting. We assume that the gravitational 
potential of the primary lens is sufficiently smooth on the 
scale of the Einstein radius of the primary lens and the 
fluxes of lensed images are perturbed by a pertuber whose
gravitational potential varies on scales smaller than 
the Einstein radius. We do not use observed fluxes as they 
would significantly distort the 
unperturbed gravitational potential. For perturbed models, however,  
we use the MIR flux ratios of lensed images as well as the 
observed positions. For simplicity, we 
assume that object X and a possible perturber reside at the lens plane.

We assume weak priors for the ellipticity $e$ and the external shear
$\gamma$ for the
primary lens galaxy G such that
$e=0.2\pm 0.2$ and $\gamma=0.1 \pm 0.1$. The expected value of 
ellipticity $e$ is taken from the elliptical isophotes of HST images 
\citep{falco1997}. The mean and 1-$\sigma$ error of external shear
$\gamma$ is obtained from quadruple lenses observed at the MIR band 
\citep{takahashi-inoue2014}. 

For the unperturbed model, we find that $\chi^2$ for the
images and the primary lens positions is extremely good as
$\chi^2_{\tr{pos}}=0.006$ (table 2). This is not surprising 
as the degree of freedom is zero if one does not impose the weak priors
on the ellipticity and external shear\footnote{The small
$\chi^2_{\tr{pos}}$ may suggest that the actual
errors of obtained positions are much smaller than the estimated values.}. 
However, $\chi^2$ for the fluxes are poor as $\chi^2_{\tr{flux}}=37.2$.
The best-fit flux ratio A2/A1$=1.038$ is consistent with the prediction of 
the fold-caustic relation A2/A1 $=1$ with an error of 
$\sim 4\%$, but it is deviated from the 
observed value A2/A1$=0.919\pm 0.021$ by $\sim 12\%$.

In order to reduce the number of model parameters, we adopt
the following assumptions: 1) In the subhalo model, the tidal radius of
an SIS is given by
$\theta_t=\sqrt{b_h/b_{\textrm{G}}}$, where $b_h$
and $b_{\textrm{G}}$ are the Einstein radii of 
an SIS and an SIE, respectively. This approximation can be verified if the subhalo
resides near A2 in the lens plane as the angular distance between A2 and
G is $1.234\sim 1$arcsec. 2) In the filament model, the typical deflection
angle due to a filament is assumed to be $b_f=0.003\,$arcsec and
the filament is perpendicular to the line-of-sight.
3) In the void model, we assume that $R_-=b_{\textrm{G}}$ and $R_+=1.3\,b_{\textrm{G}}$.
This assumption can be verified as follows. If the radius of a void is significantly larger than $b_{\textrm{G}}$,
the density perturbation mainly contributes to the constant
convergence and shear of the unperturbed lens. If smaller than $b_{\textrm{G}}$,
the contribution of line-of-sight structures to convergence is expected to
decrease\citep{inoue-takahashi2012} in comparison with larger
structures. The width of a wall of stacked voids with a radius $\sim
20-50\,h^{-1}$Mpc is expected be equal to or 
less than the radius \citep{sutter2014}. However, on much smaller scales
($\lesssim 5-10 \,h^{-1}$kpc), there have been no results yet. 
As a working hypothesis, we assume that the width of the wall 
is one third of the void radius, i.e., $d=0.3$. 
Although our assumption seems too restrictive, 
we can construct similar models with a different width of wall by
a scale transformation provided that a lensed image
lies near the edge of the wall and the width is sufficiently 
thin, i.e., $\tilde{d} \ll 1$. Suppose that a lensed image X lies
at a distance $1+d-\varepsilon$, $\varepsilon>0$ from the center of a void.
Assuming $\varepsilon \ll d\ll 1 $, equation (\ref{eq:void}) gives 
the approximated deflection angle 
$\hat{\bvec{\alpha}}_v \approx (\theta_\varepsilon \tilde{X}, \theta_\varepsilon \tilde{Y})$
, where
\BE
\theta_\varepsilon = \frac{-2 \sqrt{2}b_v \varepsilon^{3/2}}{3
d}+O(\varepsilon^{5/2})+\cdots 
\EE
Thus, at the lowest order in $\varepsilon$, 
the deflection angle $\hat{\bvec{\alpha}}_v$ is invariant 
under a scale transformation $d\rightarrow \lambda d$ and
$b_v \rightarrow \lambda b_v$, where $\lambda$ is a constant. 

\begin{figure}
\hspace{-0.3cm}
\includegraphics[width=90mm]{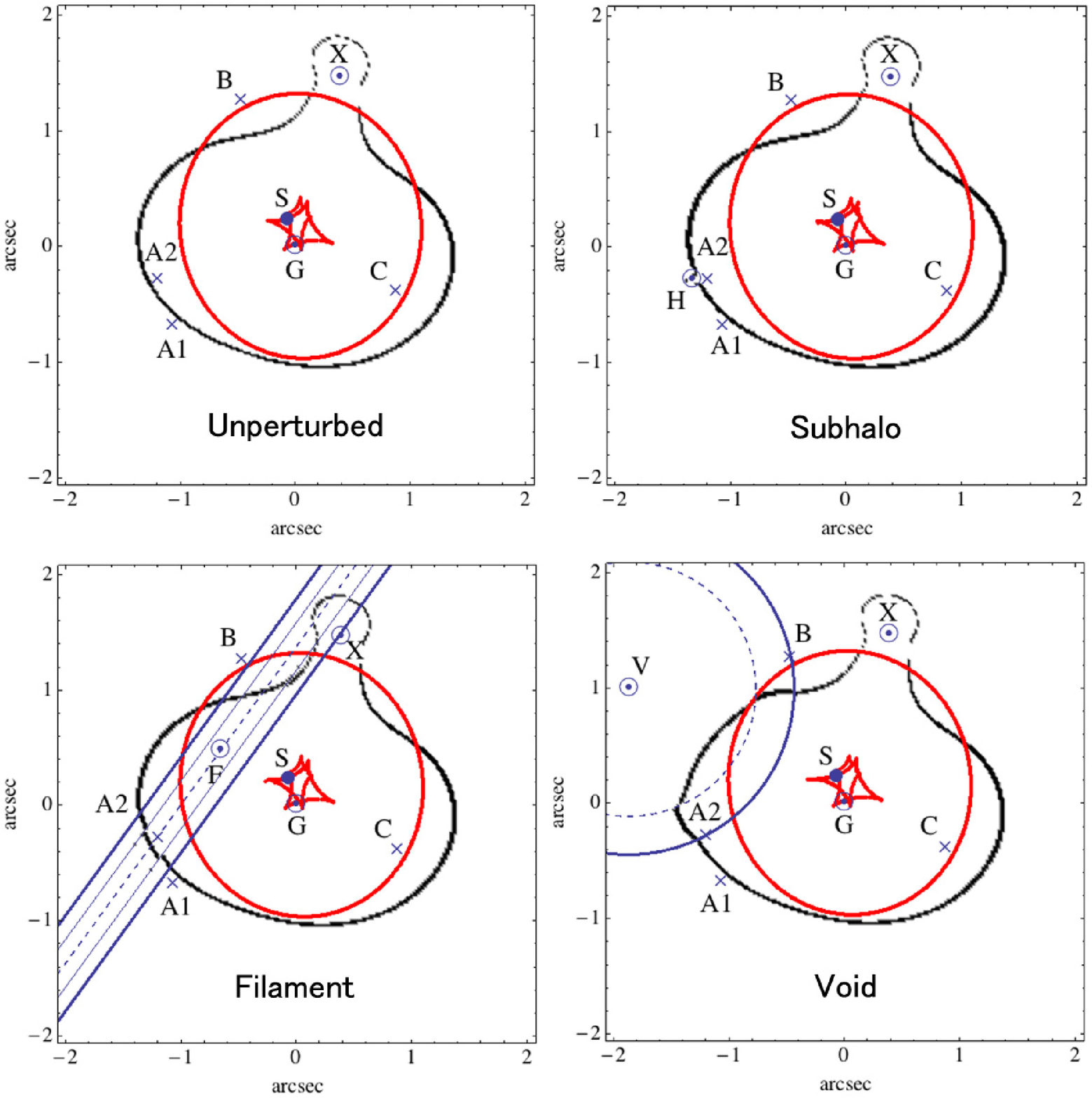}
\caption{Caustics (red) and critical curves (black) for the unperturbed
 and perturbed best-fit models. 
H, F, and V correspond to the centers of a subhalo, a 
filament, and a void, respectively. S is the position of a point
 source in each best-fit model. 
G denotes the position of the observed centroid of the primary
 lensing galaxy. X represents the position of the centroid of object X,
possibly a satellite galaxy. The central axis of the 
filament is shown by a dashed line. The boundaries of positive 
and negative density regions for the filament(void) model 
are denoted by thin(thick) and thick(thin) blue lines(curves).  }
\label{caustics}
\end{figure}

As shown in table \ref{table2}, we find that the $\chi^2$ fit for
our compensated void model ($\chi^2_{\tr{tot}}/$dof$\,=\!1.5$) 
and filament model ($\chi^2_{\tr{tot}}/$dof$\,=\!2.6$) are better than 
that for the subhalo model ($\chi^2_{\tr{tot}}/$dof$\,=\!2.8$). 
This result suggests that our void and filament
models may explain the observed anomalies in MG0514+0534
as well as the subhalo model. As shown in Fig. \ref{caustics}, we find that A2 is close to the filament and the wall of the void in the best-fit models. This suggests that the flux-ratio anomaly may be caused by a positive density perturbation in the neighbourhood of A2. 

\begin{figure}
\hspace{-0.7cm}
\includegraphics[width=95mm]{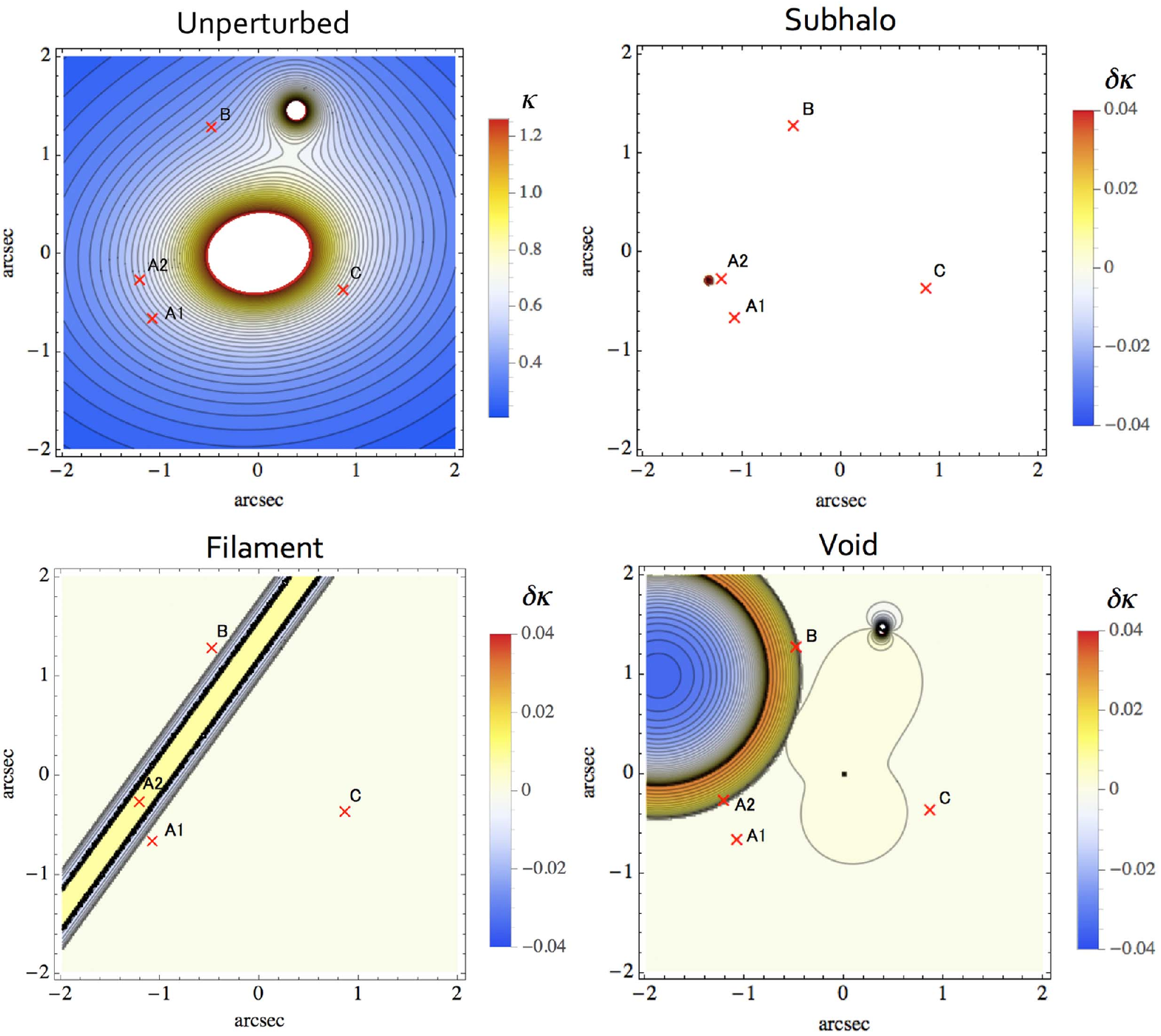}
\caption{Convergence $\kappa$ in the unperturbed model 
and the perturbation $\delta \kappa$ in the best-fit 
perturbed models. The blank regions correspond to values 
outside the range shown in each accompanying legend.}
\label{deltakappa}
\vspace*{0.3cm}
\end{figure}

We find that the truncated angular radius of the tSIS (corresponding to
$R_t$) in the best-fit model is $0.048$ arcsec 
and the mean perturbation of convergence within $R_t$ is 0.048 
(Fig. \ref{deltakappa}). The mass 
inside $R_t$ is $7\times 10^8\, h^{-1} \ms$. The angular distance
between the center of the tSIS and A2 is $d(\tr{A2})=0.13$ arcsec. 
In the best-fit filament model, it turns out that 
the surface density and the line density of the positive density 
region are $8\times 10^8\, h^{-1} \ms/\tr{arcsec}^2$ and
$2\times 10^8\, h^{-1} \ms/\tr{arcsec}$, respectively. 
The perturbation of convergence at A2 is $\delta \kappa \approx 0.008$. 
In the best-fit void model, the mass deficit is turned out to be
 $5\times 10^9\, h^{-1} \ms$ and the 
surface densities of negative and positive density regions are
$-4 \times 10^9\, h^{-1} \ms/\tr{arcsec}^2$ and
$6\times 10^9\, h^{-1} \ms/\tr{arcsec}^2$, respectively. 
The perturbation of convergence at A2 is $\delta \kappa \approx 0.004$.
Thus, the order of the surface density of a possible filament/void is 
$10^{8-9}\, h^{-1} \ms/\tr{arcsec}^2$, which can be called
``ministructures''. Within $b_{\tr{G}}$, the corresponding 
typical mass scale is $10^{8-9}\, h^{-1} \ms$. 
 
One might notice that the amplitude of the 
obtained surface density ($\sim 10^4$ times the mean cosmological mass 
density) is too big in the negative density regions.
However, we need to remember that the obtained surface mass density
corresponds to the sum of contributions from the primary lens
and the line-of-sight structures, which cannot be separated
observationally. Therefore, if some portion of 
a constant convergence is subtracted from the best-fit 
SIE-ES model, and added to the line-of-sight perturbers,
such negative density regions can be compensated by positive mass.
\begin{table*}
\hspace{-5.6cm}
\begin{minipage}{127mm}
\caption{Best-fit model parameters for MG0414+0534}
\label{symbols}
\begin{tabular}{@{}lcccc}
\hline
\hline
Model & Unperturbed & Subhalo & Filament & Void \\
\hline
 $b_{\textrm{G}}('')$ & 1.103$\pm 0.001$ & 1.103$\pm 0.001$& 1.103$\pm 0.001$ & 1.1033$\pm 0.0008$   \\
\hline
$(x_s,y_s)('')$ &  (-0.0696,0.2371) &(-0.0690,0.2372)& (-0.0694,0.2370) &
	(-0.0695,0.2368) \\
\hline
$e(\tr{G})$ & 0.231$\pm 0.001$ & 0.231$\pm 0.002$ & 0.231$\pm 0.001$ &0.231$\pm 0.001$ \\
\hline
$\phi_{e(\tr{G})}$(deg)&  -82.2$\pm 0.1$ & -82.22$\pm 0.05$ 
& -82.2$\pm0.1$ &-82.2$\pm 0.1$ \\
\hline
$\gamma$ & 0.1010$\pm 0.0004$ & 0.1010$\pm 0.0002 $ & 0.1010$\pm 0.0004$
	 &0.1010 $\pm 0.0003$ \\
\hline
$\phi_\gamma$(deg) & 53.9$\pm 0.1$ & 53.93$\pm 0.05$ & 53.9 $\pm 0.1$& 53.9 $\pm 0.1$ \\
\hline
$(x_{\tr{G}}, y_{\tr{G}})('')$ & (0.0000, -0.0001) &(0.0007,0.0000) 
 &(0.0000,-0.0001) &(0.0002, -0.0002)   \\
\hline
$b_\textrm{X}('')$ &0.184$\pm 0.001$ &0.184$\pm 0.003$ &0.184$\pm 0.001$  
&0.184$\pm 0.001$   \\
\hline
$(x_{\tr{X}}, y_{\tr{X}})('')$ & (0.383$\pm 0.008$, 1.457$\pm 0.008$) &
(0.383$\pm 0.009$,1.457$\pm 0.009$)
 &(0.383$\pm 0.008$,1.457$\pm 0.008$) &(0.383$\pm 0.008$, 1.456$\pm 0.008$)   \\
\hline
$b_h('')$ & &0.002537$\pm0.000008 $ & &   \\
\hline
$(x_h, y_h)('')$ &  &(-1.34$\pm 0.02$, -0.29$\pm 0.02$)
 &  & \\
\hline
$b_f('')$ & & &0.003(fixed) &   \\
\hline
$(x_f, y_f)('')$ &  &
 &(-0.66$\pm 0.02$,0.48$\pm 0.02$) & \\
\hline
$b_v('')$ & & & & 0.035$\pm 0.002$   \\
\hline
$(x_v, y_v)('')$ &  &
& &(-1.869$\pm 0.006$,0.990$\pm 0.003$)  \\
\hline
$R_+('')$ & & & $0.12^{+0.03}_{-0.02}$
& $1.3~b_{\textrm{G}}$(fixed)   \\
\hline
$R_-('')$ & & & $R_+$(fixed)
& $b_{\textrm{G}}$(fixed)   \\
\hline
$\chi^2_{\tr{pos}}$  &0.006 & 0.68 &0.01  & 0.07  \\
\hline
 $\chi^2_{\tr{flux}}$  &37.2 &4.87  &5.09  & 2.96 \\
\hline
$\chi^2_{\tr{weak}}$  &0.024  &0.024  &0.024 & 0.024  \\
\hline
$\chi^2_{\tr{tot}}/$dof  &37/2 &5.6/2 & 5.1/2 &3.0/2  \\
\hline
A2/A1 & 1.038&0.926 &0.920 &0.920
\\
\hline
B/A1 &0.328 &0.334 &0.328 & 0.347
\\
\hline
\label{table2}
\end{tabular}
\\
\\
Note:
$\chi^2_{\tr{tot}}$ is the sum of contributions from the
image and lens positions $(\chi^2_{\tr{pos}})$, the flux ratios
$(\chi^2_{\tr{flux}})$, and the weak priors $(\chi^2_{\tr{weak}})$ 
on $e(\tr{G})$ and $\gamma$. The parameter uncertainties were estimated 
from the range over which $\Delta \chi^2_{\tr{tot}} \le 1$ as each parameter 
was varied and the other parameters were fixed at optimised values 
except for the source position. The uncertainties of the primary lens
positions are not listed as they are well below the numerical accuracy. 
\medskip
\end{minipage}
\vspace{1cm}
\end{table*}
The three models show different patterns of convergence perturbation at
the scale $b_\tr{G}$ of the primary lens. In other
words, it implies that we can break the degeneracy of these models by
looking at global patterns of gravitational perturbations to the primary lens.

\begin{figure}
\hspace{-0.65cm}
\includegraphics[width=98mm]{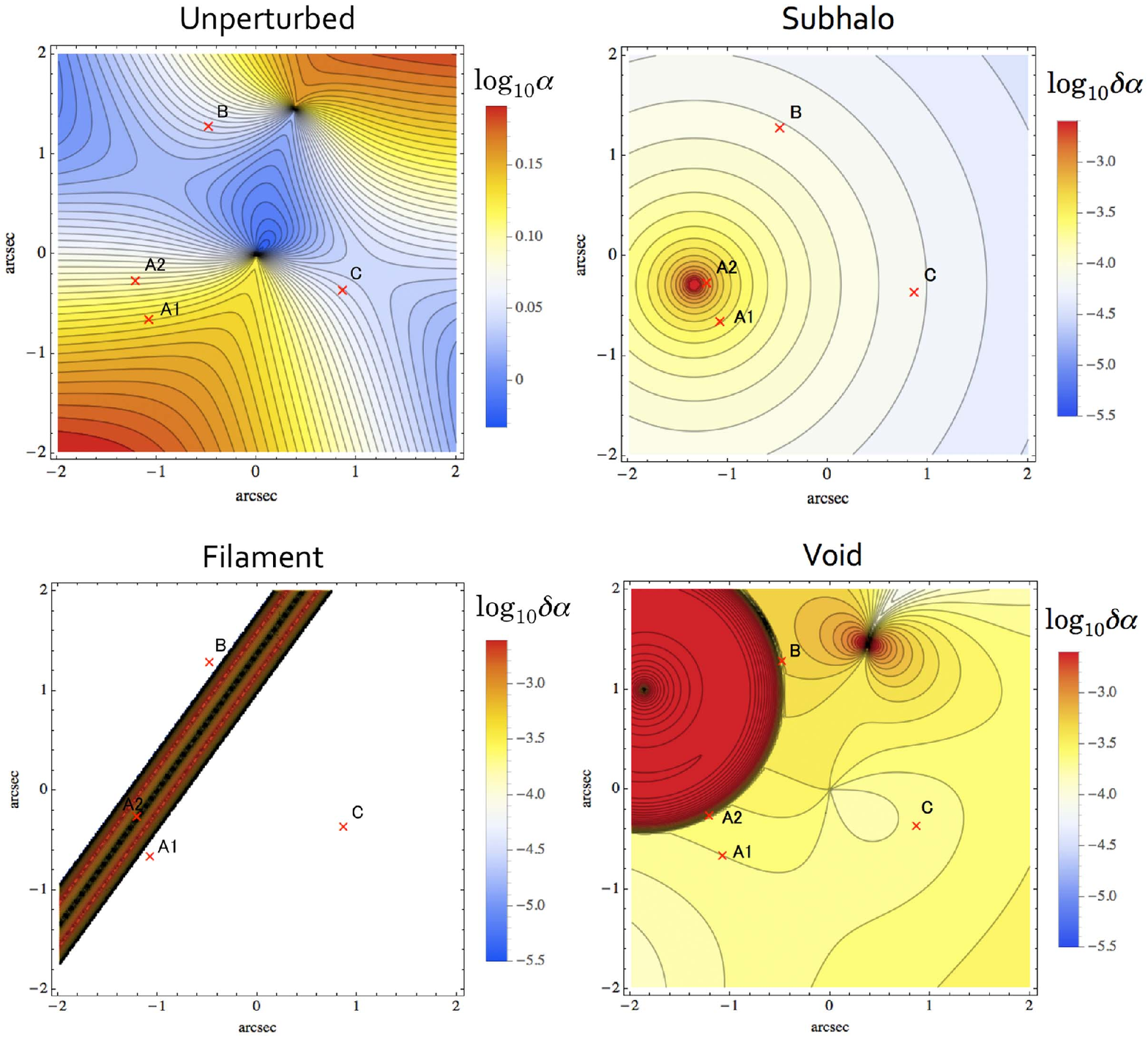}
\caption{Amplitude of the deflection angle $\alpha $ (arcsec) in the unperturbed model 
and the perturbation $\delta \alpha$ (arcsec) in the best-fit perturbed 
models. The blank regions in the lower left panel correspond to values 
outside the range shown in the accompanying legend.}
\label{deltaalpha}
\vspace*{0.3cm}
\end{figure}

The typical amplitudes of perturbations in deflection angles
at the places of the lensed images are small as $\delta \alpha \lesssim
0.001\,\textrm{arcsec}$\,(Fig. \ref{deltaalpha}). Therefore, 
the observational errors in positions of optical images
$(O(10^{-3})\,\textrm{arcsec})$ still admit 
a variety of lens models that predict similar flux ratios
within the relative errors of $\lesssim 0.01$. 

In order to check this point, we calculate the magnification perturbations
caused by perturbers. 
Indeed, we find similar magnification perturbations
at the place of A2 for all the three models. However, the patterns of 
magnification perturbations turn out to be different at other places. 
This is due to the difference in convergence and shear on scales
$\lesssim b_\tr{G}$. Therefore, we may be able to break the degeneracy 
by measuring the differential magnifications, which will be discussed in
the next section.
\begin{figure}
\hspace{-0.45cm}
\includegraphics[width=95mm]{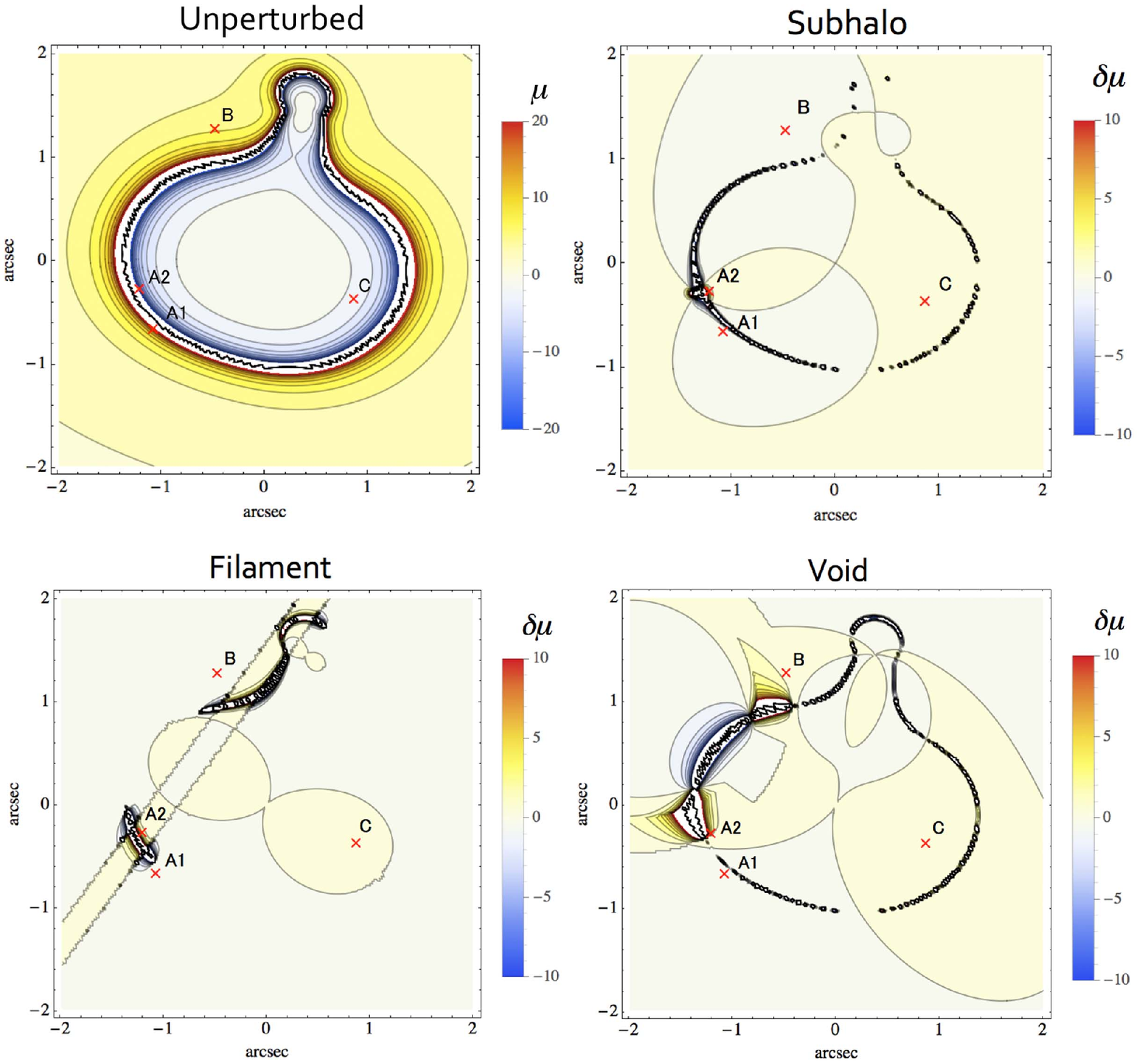}
\caption{Magnification $\mu$  in the unperturbed model and the
 perturbations $\delta \mu$ in the best-fit perturbed models. 
The blank regions correspond to values 
outside the range shown in each accompanying legend. }
\label{deltamu}
\vspace*{0.3cm}
\end{figure}
\vspace{3cm}
\section{Differential magnification effect}
\begin{figure}
\includegraphics[width=85mm]{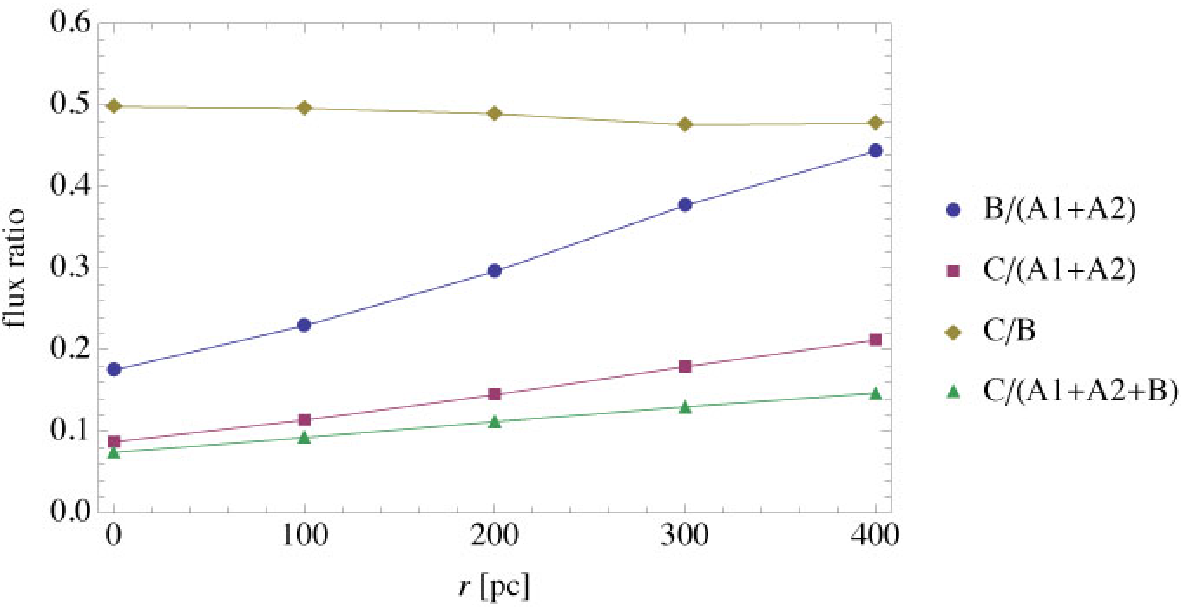}
\includegraphics[width=85mm]{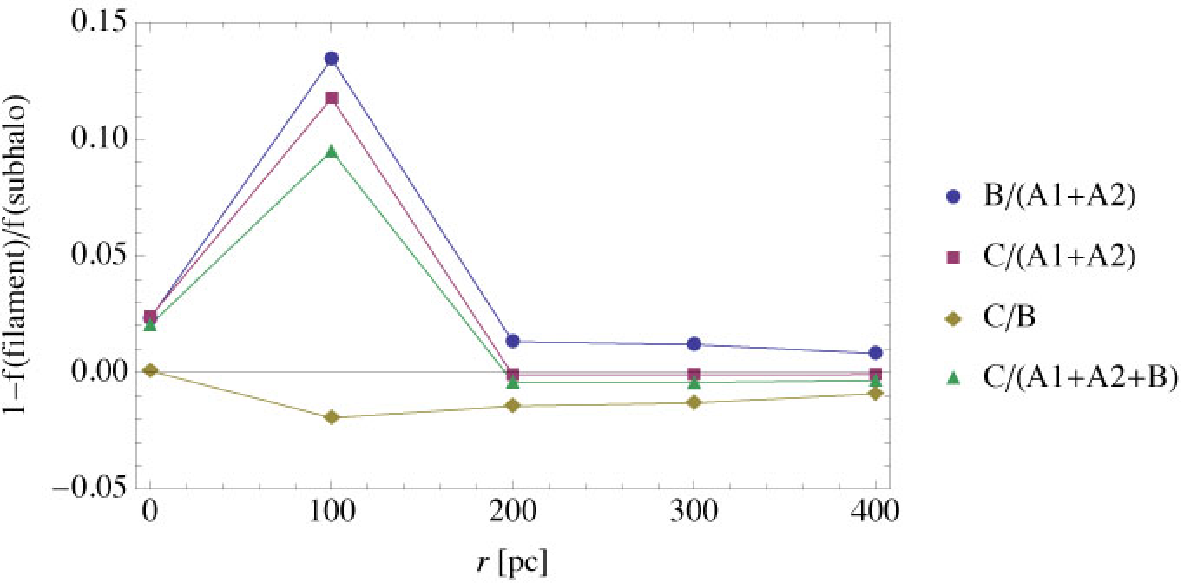}
\includegraphics[width=85mm]{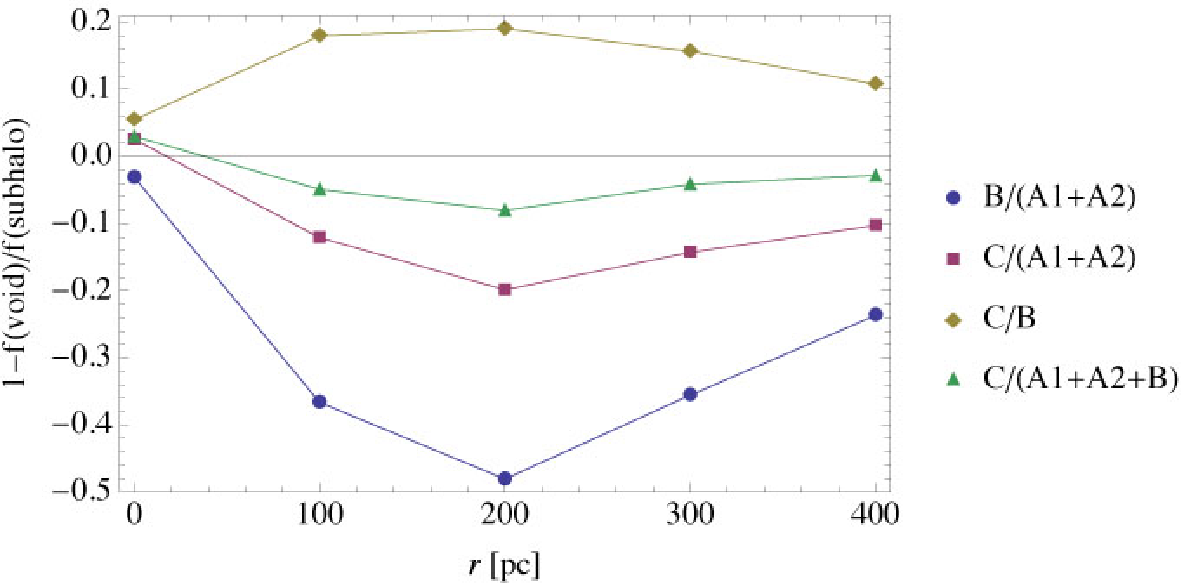}
\caption{Flux ratios of the subhalo model (top), the
 relative differences between the filament and subhalo models (middle) and
 those between the void and subhalo models (bottom) as a function of the 
source radius $r$. }
\label{flux-ratios}
\vspace*{0.3cm}
\end{figure}

As shown in Fig. \ref{deltamu}, the perturbation of 
magnification at the positions of lensed images are similar 
in all the models. However, the global pattern of perturbation
differs in each case. In order to discriminate the three models,
we consider the differential magnification effect. For simplicity, as
the surface brightness of an extended source, we
assume a circular Gaussian profile centred at the best-fit position of a
point source. The source radius $r$ is represented by the standard deviation
$\sigma$. As one can see in Fig. \ref{flux-ratios}, the flux ratios
vary as a function of a source radius $r$.  The flux ratios of 
the void model and the subhalo model differ by $5\sim 50$\% for 
$0<r<400$\,pc. On the other hand, the difference between the filament model
and the subhalo model is only $1 \sim 2$\% except for $r\sim
100\,$pc (see Fig. \ref{flux-ratios}). This is because that the pattern of magnification perturbations
of the filament and the subhalo model are similar in the neighbourhood
of A1 and A2. However, if the position of the extended source is not centred at 
the position of the best-fit point source, we may detect a noticeable
change if the lensed image in the neighbourhood of B crosses
the filament. The red-blue decomposition of line emission may work
for this purpose.
  
With errors of the order of
$<0.1$ per cent in the flux ratios and $<0.1$ arcsec in the 
angular resolution, we can discriminate the possible models that can
account for the anomaly in the flux ratios. Thus, 
observation of continuum and line emissions of MG0414+0534 
in the submillimeter bands with ALMA (Atacama Large
Millimeter/submillimeter Array) is of great importance to probe the
origin of the flux ratio anomalies. Multifrequency observation 
may be necessary to break the model degeneracy.
\vspace{0cm}
\section{Conclusion and Discussion}
In this paper, we have found 
that the observed flux-ratio anomaly in MG0414+0534 
can be explained by a presence of either 
a minifilament or a minivoid in the line-of-sight with a 
surface mass density of the order of 
$10^{8-9}\,h^{-1}\ms/\tr{arcsec}^2$ without taking into
account any subhalos in the lensing galaxy. The astrometric perturbation  
by a possible minifilament/minivoid is $\lesssim
 0.001\,\tr{arcsec}$ and the amplitudes of convergence perturbations 
due to these perturbers are 
typically $\kappa \sim 0.004-0.008$ at the place of an 
image A2 that shows anomaly in the flux. In order to discriminate
models with the line-of-sight ministructures 
from those with a subhalo(s) in the lensing galaxy, 
the differential magnification effect can be used
if the source size is $>100\,\tr{pc}$. 

In our analysis, we have used very simple models for
a filament and void. They are assumed to be locally homogeneous and 
residing in the lens plane. We have found that all the best-fit
models show that image A2 is perturbed by a perturber with a 
positive convergence. Although it is very important to study
more realistic mass distributions for modeling 
the filament/void, the mass scale of possible 
perturber would not change much. It is also important 
to consider the cases in which multiple perturbers in the line-of-sight
affect the lensed images. In order to do so, we need
to carry out more realistic numerical simulations
in which these perturbers reside at arbitrary 
places in the line-of-sight. 

In order to break the degeneracy of possible models, we
need observation at the submillimeter band using 
interferometers such as ALMA. If the perturber is massive enough, 
we may be able to directly map the gravitational perturbation 
by lensing \citep{inoue-chiba2003, inoue-chiba2005}. By observing continuum and 
line emissions from circum-nuclear dusts around quasars 
at multi-frequency sub-millimeter bands, we will be able to put 
stringent constraints on possible models of perturbers.

Another possible way to break the model degeneracy is to use
time delay of optical or near-infrared lensed images. Although
the convergence perturbation around A2 are expected to be similar,  
the gravitational potential projected along the line-of-sight 
can be different in principle \citep{mccully2014,schneider2014}. 
We expect that the perturbation of 
time delay is conspicuous for matter distributions that are
locally over-compensating since the potential wells are deep. 
This issue will be analyzed in detail in our future work.

If the anomaly is related to ministructures in the line-of-sight
rather than subhalos in the lensing galaxy, then
we expect a correlation between perturbed fluxes of lensed images.
If it is caused by subhalos, such a spatial 
correlation is not expected as they reside 
randomly in galactic halos. It is very important to assess
the number of lens systems necessary for making such distinction.

Minifilaments and walls of minivoids may retain plenty of 
HI gas. Then HI 21-cm absorption systems in the line-of-sight 
would have a correlation with magnification perturbations 
as well as the optical-near-infrared colour due to the presence of dust.
The observed HI 21-cm absorption systems at $z=0.3-0.5$ in the line-of-sight 
of MG0414+0534 \citep{tanna2013} may be connected with such
minifilaments or walls. Thus, correlation study between reddening,
absorption lines, and flux-ratio anomalies will be much important in near future
in order to understand the clustering property of dark matter and
baryons at $\sim 10\,$kpc scales.

\section{Acknowledgments}
The author thanks Masashi Chiba and Ryuichi Takahashi for
valuable discussion and useful comments. 
This work is supported in part by JSPS Grant-in-Aid for
Scientific Research (B) (No. 25287062) ``Probing the origin
of primordial minihalos via gravitational lensing phenomena''.
\bibliographystyle{mn2e}
\bibliography{weak-lensing-by-los2014}

\end{document}